\documentclass[12pt,letterpaper]{article}

\usepackage[letterpaper,margin=1in]{geometry}

\usepackage{etoolbox}
\newcommand{\spacingset}[1]{%
  \renewcommand{\baselinestretch}{#1}\small\normalsize}
\AtBeginDocument{%
  \spacingset{1.45}%
  \captionsetup{font=normalsize}%
  \setcitestyle{round,semicolon,aysep={},yysep={,}}%
  \setlength{\bibsep}{0pt}%
}

\usepackage{amsmath,amssymb,amsthm,bm,mathtools}
\usepackage{booktabs,array,multirow,longtable}
\usepackage{graphicx}
\usepackage{caption}
\usepackage{subcaption}
\usepackage{pdflscape}
\usepackage{enumitem}
\usepackage{needspace}
\usepackage[round,authoryear]{natbib}
\usepackage[ruled,vlined,linesnumbered]{algorithm2e}
\usepackage{float}
\usepackage{placeins}
\usepackage[hidelinks]{hyperref}
\usepackage[nameinlink,capitalize]{cleveref}
\usepackage{microtype}

\allowdisplaybreaks
\AtBeginEnvironment{tabular}{}
\newtheorem{theorem}{Theorem}
\newtheorem{proposition}{Proposition}

\newtheorem{corollary}{Corollary}
\theoremstyle{definition}

\theoremstyle{remark}

\SetKwInOut{Input}{Input}
\SetKwInOut{Initialize}{Initialize}
\SetKwInOut{Output}{Output}
\SetAlgoNlRelativeSize{0}
\SetAlFnt{\normalsize\spacingset{1.45}}

\newcommand{\Prob}{\mathbb P}

\title{\bfseries
TRACE: Adaptive Credibility Reweighting for Robust Bayesian Model Selection with Source Summary Statistics
}
\date{}

\begin{document}
\spacingset{1}

\author{Pengfei Chen\thanks{%
All four authors are affiliated with the Center for Mathematics and Interdisciplinary Sciences,
Fudan University, Shanghai 200433, China, and the Shanghai Institute for Mathematics and
Interdisciplinary Sciences (SIMIS), Shanghai 200433, China.
E-mail: \href{mailto:24114020003@m.fudan.edu.cn}{24114020003@m.fudan.edu.cn} (Pengfei Chen),
\href{mailto:24114020017@m.fudan.edu.cn}{24114020017@m.fudan.edu.cn} (Yuehan Yang),
\href{mailto:lucyk@fudan.edu.cn}{lucyk@fudan.edu.cn} (Lucy Klinger), and
\href{mailto:chongy@fudan.edu.cn}{chongy@fudan.edu.cn} (Chong You).
Corresponding authors: Lucy Klinger and Chong You.}%
\quad Yuehan Yang\quad Lucy Klinger\quad Chong You}
\maketitle


\thispagestyle{plain}

\begin{abstract}
This study focuses on regression estimation and variable selection based on target data, while incorporating information from coefficient estimates obtained in related studies. A weighted combination of these estimates may agree with some target effects but differ from others. We propose Transfer Regression with Adaptive Credibility Estimation (TRACE), a Bayesian method requiring only fitted coefficients and sample sizes from source studies. TRACE learns simplex-constrained source weights shared across variables. Separate latent indicators determine target inclusion and whether each coefficient is guided by the weighted source estimate. We develop an approximate collapsed variational expectation-maximization algorithm with robust source-weight initialization. Under suitable conditions, consistency of variable selection, transfer decisions, and source-weight estimation is established. These results hold at every prespecified fixed iteration in fixed dimension and at global mean-field optima of the exact collapsed variational objective as the dimension grows. Simulations show that TRACE achieves competitive variable-selection performance under different patterns of source heterogeneity. An application to data from the Genotype-Tissue Expression project illustrates its use in cross-tissue gene-expression prediction.
\end{abstract}


\vfill
\newpage
\spacingset{1.45}
\section{Introduction}
\label{sec:introduction}

A regression model developed in one population may contain useful information about
another without being directly applicable to it. Transfer learning seeks to 
use information from related source studies to improve estimation, prediction, 
and variable selection in a target population \citep{li2022transfer,suder2025bayesian}. 
The challenge is that source--target similarity may vary across covariates: 
a source can provide useful information about some target effects while being misleading 
about others. Whether a predictor is relevant to the target response is therefore distinct 
from whether its source information should be borrowed. We consider this problem when 
individual-level target observations are available, but each source study provides only 
fitted regression coefficients and its sample size. Such information may remain available 
when data-use agreements, privacy concerns, or legacy analyses preclude access to the original source records. The task is then to combine the available source estimates and determine where source information can improve target estimation and variable selection, without access to individual-level source data.

Statistical approaches to high-dimensional transfer learning have established estimation, 
prediction, and inference guarantees by relating source and target regression parameters. 
Important developments include transfer learning for sparse linear regression 
\citep{li2022transfer}, extensions to generalized linear models \citep{tian2023glm,li2024knowledge}, and methods for federated analysis across heterogeneous 
populations \citep{li2023federated}. A related literature studies how individual-level data 
from an internal study can be combined with external summaries or previously fitted models. 
Constrained estimation \citep{chatterjee2016constrained} and synthetic-data integration 
\citep{gu2023summary} provide ways to incorporate such information without directly pooling 
the original records. More recently, \citet{gu2025angle} developed angle-based transfer learning using parameter estimates from pretrained source models. These contributions establish that the information available from a source, as well as its relationship to the target, is central to the design of a transfer procedure.

Bayesian methods offer a complementary approach by representing external information through 
priors and learning its influence from target observations; \citet{suder2025bayesian} review 
this literature. Recent developments include random-effect calibration of source predictions 
\citep{hickey2024recast} and shrinkage formulations for differences between source and target 
parameters \citep{abba2026shrinkage}. Two methods are particularly relevant here. CONCERT 
uses a conditional spike-and-slab construction to represent covariate-specific similarities 
between target and source parameters, allowing information transfer and variable selection 
in the target model \citep{zhang2024concert}. TRADER uses a guided horseshoe prior centered 
at an adaptively weighted combination of source estimates, with local shrinkage controlling 
information transfer for each coefficient \citep{lai2026bayesian}. These approaches provide 
different ways to use heterogeneous external information. They motivate studying how to combine source estimates, select target variables, and decide where to borrow from the combined estimate when only coefficient summaries are available. Section~\ref{sec:related-methods} compares these methods and discusses the specific problems addressed in this article.

Posterior computation is central to Bayesian variable selection. Variational Bayes (VB) 
approximates a posterior by optimizing over a tractable family of distributions, 
but mean-field approximations can underestimate posterior uncertainty, and the corresponding 
coordinate-ascent algorithms can be sensitive to initialization \citep{blei2017variational}. 
In regression variable selection, \citet{ormerod2017variational} develop an initialization and tuning strategy alongside their consistency analysis. Collapsed variational Bayes (CVB) 
integrates out a subset of parameters before approximating the remaining posterior 
\citep{teh2007collapsed}. This accounts analytically for uncertainty in these parameters and 
avoids the additional approximation error introduced by specifying separate mean-field factors for them. For linear regression, \citet{you2023ollapsed} develop an approximated CVB method with selection consistency results and report performance comparable to a path-search-based VB algorithm at substantially lower computational cost. In high dimensions, 
\citet{ray2022variational} establish posterior contraction rates for mean-field approximations under sparse priors. These developments motivate extending CVB to transfer learning, where variable selection, transfer decisions, and source weights must be determined together.

In this study, we propose Transfer Regression with Adaptive Credibility Estimation(TRACE), a Bayesian procedure for regression estimation and variable selection using source coefficient summaries. TRACE combines source estimates through simplex-constrained weights and uses separate latent indicators for target inclusion and coordinate-specific information transfer. An approximated CVB procedure integrates out regression coefficients and noise variance before jointly estimating source weights and indicator probabilities. This permits target selection and selective transfer without individual-level source observations.

The theory in this study covers both fixed and growing dimensions. In fixed dimension, we establish selection consistency for the target variables and transfer indicators, together with consistency of the estimated source weights, for any fixed number of iterations. When the dimension grows, we establish consistency of the estimated source weights, selection consistency for the target variables, and consistency of transfer decisions at the global variational optimum.

Simulation studies evaluate target selection, transfer decisions, and prediction under heterogeneous source configurations. An application to cross-tissue gene-expression prediction using GTEx data illustrates the method's practical use.

The remainder of the article is organized as follows. Section~\ref{sec:related-methods} reviews related methods and the specific unresolved issues. Section~\ref{sec:methodology} presents the model, priors, and CVB algorithm. Section~\ref{sec:theory} develops the theoretical results. 
Sections~\ref{sec:simulation} and~\ref{sec:real-data} report simulations and the real-data 
application, respectively. Section~\ref{sec:conclusion} concludes with a discussion of limitations and future directions. Proofs, detailed derivations, and additional numerical results are provided in the Supplementary Material.

\section{Related Methods and Problems}
\label{sec:related-methods}

Existing transfer-learning methods differ in the source information they require
and in how source information is incorporated into the target analysis. These
differences are particularly relevant to TRACE, where only fitted source and sample sizes
coefficients are available and the target analysis must determine both how the
sources should be combined and where their information should be used.

\subsection{Source data and coefficient summaries}
\label{subsec:related-source-information}

Trans-Lasso \citep{li2022transfer} requires individual-level source data and
uses the source and target samples jointly for transfer estimation. CONCERT
\citep{zhang2024concert} jointly models source-specific and target regressions
and uses the corresponding source likelihood and design information. This
allows source-specific regression structures and covariate-specific similarities
with the target to be modeled directly. In Gaussian linear regression, the
required source information can be represented by sufficient statistics, so
individual source records need not be retained once these quantities are
available. Nevertheless, CONCERT requires more source information than fitted
regression coefficients alone.

TRADER \citep{lai2026bayesian} uses coefficient estimates from pre-trained
source models without requiring individual-level source data or source design
matrices. TRACE uses fitted regression coefficients and source sample sizes to set 
its borrowing precision. from multiple source studies. The main difference between TRADER
and TRACE therefore lies in how these source estimates are combined and how
their influence on individual target variables is controlled.

\subsection{Variable selection, source weighting, and information transfer}
\label{subsec:related-selection-borrowing}

Target variable selection, source weighting, and coordinate-specific information transfer play different
roles in transfer learning. Variable selection determines whether a predictor
is relevant to the target response. Source weighting determines how information
from multiple source studies is combined. A transfer decision determines whether the
resulting source information should influence the target coefficient for a
particular variable. A predictor may be relevant to the target even when its
source effects differ substantially from the target effect, and a source
combination that is useful for some variables may be inappropriate for others.

CONCERT separates target variable selection from source--target similarity.
Its similarity indicators are specific to both the source and the variable,
allowing information transfer to vary across sources and variables. Thus, source
compatibility is assessed before the source information is reduced to a common
coefficient estimate.
TRADER takes a different approach. It constructs an adaptively weighted
combination of the fitted source coefficients and uses this combination as the
center of a guided horseshoe prior. Coefficient-specific local shrinkage then
determines the extent to which each target coefficient is pulled toward the
combined source estimate. Hence, source weighting and variable-specific information transfer are both adaptive, with the latter represented through continuous
shrinkage.
In TRACE, the source coefficient estimates are combined through
simplex-constrained weights, while the use of source information is determined separately for each
variable. The source weights are shared across variables and determine the
relative contribution of each source study to the common source estimate. A
target inclusion indicator determines whether a variable enters the target
model, whereas a transfer indicator determines whether its coefficient is
guided by the resulting weighted source estimate. TRACE therefore assigns
separate probabilities to target inclusion and information transfer.

Source weighting and variable-specific information transfer address different forms of
heterogeneity. Source studies may differ in their relative contribution to a
useful common source estimate, while the resulting weighted estimate may still
be incompatible with the target for some variables. Source weighting determines
the common combination, whereas the transfer indicators determine where that combination is
used. The two components are therefore distinct but related: the estimated
source weights affect the source information considered for transfer, while
the variables for which information transfer is appropriate provide information about the
source weights. How TRACE models and estimates these components is described in
Section~\ref{sec:methodology}.

\section{Methodology}
\label{sec:methodology}

\subsection{Model and priors}
\label{subsec:hierarchical-main}

Let $(\bm X,\bm y)$ denote the target data, where $\bm X\in\mathbb R^{n\times p}$ and 
$\bm y\in\mathbb R^{n}$. Source study $k$ provides a fitted coefficient vector 
$\hat{\bm b}^{(k)}\in\mathbb R^p$ and its sample size $n_k$, for $k=1,\ldots,K$. Write
\[
 n_s=\frac1K\sum_{k=1}^K n_k,\qquad
 \bm B=(\hat{\bm b}^{(1)},\ldots,\hat{\bm b}^{(K)}) \in\mathbb R^{p\times K},
\]
and denote the $j$th row of $\bm B$ by
\[
 \hat{\bm b}_j^\top=(\hat b_j^{(1)},\ldots,\hat b_j^{(K)}).
\]
Throughout, $p\geq2$ and $\bm X$ has no zero columns.

We treat the source summaries as observed inputs and suppress conditioning on $\bm X$ and $\bm B$ 
in the densities below. Write $[p]=\{1,\ldots,p\}$. For index sets $u,v$, $\bm a_u$ denotes the 
subvector of $\bm a$ indexed by $u$, and $\bm A_{u,v}$ denotes the submatrix of $\bm A$ with rows 
in $u$ and columns in $v$. For row and column index sets $u$ and $v$ of $\bm A$, $\bm A_{u\cdot}$ 
selects the rows in $u$ and $\bm A_v$ selects the columns in $v$. For predictor index sets $v\subseteq[p]$, 
write $-v=[p]\setminus v$.

The source weights $\bm\alpha$ are shared across predictors and belong to
\[
 \Delta_K=\{\bm\alpha\in\mathbb R^K:
         \bm\alpha\geq\bm0,\ \bm1^\top\bm\alpha=1\}.
\]
They determine the relative contributions of the sources to the combined estimate $\bm B\bm\alpha$.
For each predictor, $z_j$ indicates inclusion in the target model, while $c_j$ indicates information 
transfer from the corresponding combined source estimate. Let $\bm Z=\operatorname{diag}(z_1,\ldots,z_p)$. 
We specify
\begin{equation*}
\begin{aligned}
 \bm y\mid\bm\beta,\bm Z,\sigma_y^2
 &\sim\mathcal N_{n}(\bm X\bm Z\bm\beta,\sigma_y^2\bm I_{n}),
 & z_j\mid\rho_z&\sim\operatorname{Bernoulli}(\rho_z),\\
 \beta_j\mid\bm\alpha,c_j,\sigma_y^2
 &\sim\mathcal N\!\left(c_j\bm\alpha^\top\hat{\bm b}_j,\sigma_y^2/\tau_j\right),
 & c_j\mid\rho_c&\sim\operatorname{Bernoulli}(\rho_c),\\
 \tau_j
 &=c_j(\bm T_x)_{jj}+(1-c_j)\tau_0,
 & \sigma_y^2&\sim\operatorname{Inv\text{-}Gamma}(A_y,B_y),
\end{aligned}
\end{equation*}
where $0<\rho_z,\rho_c<1$ and $\tau_0,A_y,B_y>0$ are hyperparameters, with $\tau_0$ specifying the 
baseline precision relative to $\sigma_y^2$, and
\begin{equation}
\bm T_x=\frac{n_s}{n\log p}
\operatorname{diag}(\bm X^\top\bm X).
\label{eq:prior-precisions-main}
\end{equation}

When $c_j=1$, the prior is centered at $(\bm B\bm\alpha)_j$ and becomes more concentrated as the 
source sample size increases. When $c_j=0$, the prior is centered at zero with variance $\sigma_y^2/\tau_0$.
Note that we estimate $\bm\alpha$ over $\Delta_K$ rather than assign it a prior, avoiding an additional 
integration or a variational approximation over the simplex. The indicator posterior is evaluated 
conditional on these weights.

\subsection{Collapsed variational approximation}
\label{subsec:vb-main}
Following the CVB approach of \citet{you2023ollapsed}, we integrate out $\bm\beta$ and $\sigma_y^2$ before approximating the posterior of
$(\bm z,\bm c)$. Define
\[
 \bm T=\operatorname{diag}(\tau_1,\ldots,\tau_p).
\]
The conditional prior precision matrix is $\bm T/\sigma_y^2$.
The collapsed log marginal likelihood, up to a term independent of $(\bm z,\bm c,\bm\alpha)$, is
\begin{equation}
\begin{aligned}
 \log p(\bm y\mid\bm z,\bm c,\bm\alpha)
 ={}&\frac12\log|\bm T|
     -\frac12\log|\bm T+\bm Z\bm X^\top\bm X\bm Z|\\
 &-\left(A_y+\frac{n}{2}\right)
   \log\!\left\{B_y+\frac12R(\bm z,\bm c,\bm\alpha)\right\},
\end{aligned}
\label{eq:collapsed-main}
\end{equation}
where
\begin{equation}
\begin{aligned}
 R(\bm z,\bm c,\bm\alpha)
 ={}&\bm y^\top\bm y
 +\bm\alpha^\top\bm B^\top\bm T_x\operatorname{diag}(\bm c)\bm B\bm\alpha\\
 &-\left(\bm Z\bm X^\top\bm y+\bm T_x\operatorname{diag}(\bm c)\bm B\bm\alpha\right)^\top
 \left(\bm T+\bm Z\bm X^\top\bm X\bm Z\right)^{-1}\\
 &\quad\left(\bm Z\bm X^\top\bm y+\bm T_x\operatorname{diag}(\bm c)\bm B\bm\alpha\right).
\end{aligned}
\label{eq:collapsed-residual-main}
\end{equation}
The detailed derivation is given in Section~A.1 of the Supplementary Material.

For fixed $\bm\alpha$, we approximate the marginal posterior $p(\bm z,\bm c\mid\bm y,\bm\alpha)$ 
within the mean-field family $\mathcal Q$ of Bernoulli product distributions,
\begin{equation}
 \mathcal Q
 =\left\{q: q(\bm z,\bm c)=\prod_{j=1}^p q(z_j)\prod_{j=1}^p q(c_j)\right\},\quad
 \omega_j=q(z_j=1),\quad \gamma_j=q(c_j=1),
 \label{eq:mean-field-main}
\end{equation}
where $0\leq\omega_j,\gamma_j\leq1$. The probabilities describe variable inclusion and information 
transfer, respectively. We write the optimal approximation as
\begin{equation*}
 q_{\bm\alpha}^{*}\in
 \operatorname*{arg\,min}_{q\in\mathcal Q}
 \operatorname{KL}\!\left\{q(\bm z,\bm c)\,
 \middle\|\,p(\bm z,\bm c\mid\bm y,\bm\alpha)\right\}.
\end{equation*}
Equivalently, it maximizes the evidence lower bound
\begin{equation}
 \mathcal L(q,\bm\alpha)
 =\mathbb E_q\!\left[\log\frac{p(\bm y\mid\bm z,\bm c,\bm\alpha)p(\bm z)p(\bm c)}{q(\bm z,\bm c)}\right].
 \label{eq:collapsed-elbo-main}
\end{equation}
Joint variational estimation maximizes $\mathcal L(q,\bm\alpha)$ over $q\in\mathcal Q$ and 
$\bm\alpha\in\Delta_K$. The optimal factors are Bernoulli distributions with probabilities
$\omega_j^*(\bm\alpha)=q_{\bm\alpha}^*(z_j=1)$ and $\gamma_j^*(\bm\alpha)=q_{\bm\alpha}^*(c_j=1)$ 
satisfying
\begin{align}
 \operatorname{logit}\{\omega_j^*(\bm\alpha)\}
 =&\operatorname{logit}(\rho_z)
 +\mathbb E_{q^*_{\bm\alpha,-z_j}}\!\left[
   \log\frac{p(\bm y\mid z_j=1,\bm z_{-j},\bm c,\bm\alpha)}
                 {p(\bm y\mid z_j=0,\bm z_{-j},\bm c,\bm\alpha)}
 \right],\label{eq:omega-main}\\
 \operatorname{logit}\{\gamma_j^*(\bm\alpha)\}
 =&\operatorname{logit}(\rho_c)
 +\mathbb E_{q^*_{\bm\alpha,-c_j}}\!\left[
   \log\frac{p(\bm y\mid c_j=1,\bm c_{-j},\bm z,\bm\alpha)}
                 {p(\bm y\mid c_j=0,\bm c_{-j},\bm z,\bm\alpha)}
 \right].\label{eq:gamma-main}
\end{align}
Here $q^*_{\bm\alpha,-z_j}$ and $q^*_{\bm\alpha,-c_j}$ are the products of the remaining optimal 
factors. These equations characterize the factors implicitly, since each expectation depends on 
the other factors. Their derivation is given in Section~A.2 of the Supplementary Material.

\subsection{Estimation algorithm}
\label{subsec:approx-cvb-main}
\paragraph{Approximating the expectations.}
For a given $\bm\alpha$, the equations for $\omega_j^*(\bm\alpha)$ and $\gamma_j^*(\bm\alpha)$, 
conditional on $q_{\bm\alpha,-z_j}^*$ and $q_{\bm\alpha,-c_j}^*$, involve expectations of logarithms 
and matrix inverses. Their explicit forms and derivations are given in Section~A.2 of the Supplementary 
Material. These expectations generally require summation over the remaining indicator values. 
We approximate them by substituting the corresponding matrix expectations before evaluating the nonlinear functions.
Let
\begin{equation}
\begin{aligned}
\bm W_{\bm\alpha}^{\ast}
&=
\mathbb E_{q_{\bm\alpha}^{\ast}}(\bm Z)
=
\operatorname{diag}\{\bm\omega^{\ast}(\bm\alpha)\},
\qquad
\bm\Gamma_{\bm\alpha}^{\ast}
=
\mathbb E_{q_{\bm\alpha}^{\ast}}(\operatorname{diag}(\bm c))
=
\operatorname{diag}\{\bm\gamma^{\ast}(\bm\alpha)\},
\\
\bm\Omega_{\bm\alpha}^{\ast}
&=
\mathbb E_{q_{\bm\alpha}^{\ast}}(\bm z\bm z^\top)
=
\bm\omega^{\ast}(\bm\alpha)
\bm\omega^{\ast}(\bm\alpha)^\top
+\bm W_{\bm\alpha}^{\ast}
(\bm I_p-\bm W_{\bm\alpha}^{\ast}).
\end{aligned}
\label{eq:variational-matrices}
\end{equation}
Then
\begin{equation*}
\begin{aligned}
\mathbb E_{q_{\bm\alpha}^{\ast}}
(\bm Z\bm X^\top\bm X\bm Z)
&=
\bm X^\top\bm X\odot\bm\Omega_{\bm\alpha}^{\ast},
\\
\bm T_{\bm\alpha}^{\ast}
=\mathbb E_{q_{\bm\alpha}^{\ast}}(\bm T)
&=\bm\Gamma_{\bm\alpha}^{\ast}\bm T_x
+
\tau_0(\bm I_p-\bm\Gamma_{\bm\alpha}^{\ast}),
\\
\mathbb E_{q_{\bm\alpha}^{\ast}}
(\bm T_x\operatorname{diag}(\bm c))
&=\bm\Gamma_{\bm\alpha}^{\ast}\bm T_x,
\end{aligned}
\end{equation*}
where $\odot$ denotes entrywise multiplication. For any $q\in\mathcal Q$ with the probability parameters 
$(\bm\omega,\bm\gamma)$, let $\bm W$, $\bm\Gamma$, and $\bm\Omega$ denote the corresponding expectations 
as in \eqref{eq:variational-matrices}.

For $\bm\omega\in[0,1]^p$, define $D=\{j\in[p]:\omega_j=0\}$. In $\mathbb E_q\{R(\bm z,\bm c,\bm\alpha)\}$, 
the terms involving coordinates in $D$ cancel exactly. Substituting the expectations of the remaining 
matrix quantities therefore gives the following expression on $-D$:
\begin{equation*}
\begin{aligned}
\widetilde R(\bm\omega,\bm\gamma,\bm\alpha)
={}&\bm y^\top\bm y
+\bm\alpha^\top\bm B_{-D\cdot}^\top
(\bm\Gamma\bm T_x)_{-D,-D}
\bm B_{-D\cdot}\bm\alpha
\\
&-\left\{
\bm W_{-D,-D}\bm X_{-D}^\top\bm y
+(\bm\Gamma\bm T_x)_{-D,-D}
\bm B_{-D\cdot}\bm\alpha
\right\}^\top
\\
&\quad\left\{
\left[\bm\Gamma\bm T_x+\tau_0(\bm I_p-\bm\Gamma)\right]_{-D,-D}
+\bm X_{-D}^\top\bm X_{-D}\odot\bm\Omega_{-D,-D}
\right\}^{-1}
\\
&\quad\left\{
\bm W_{-D,-D}\bm X_{-D}^\top\bm y
+(\bm\Gamma\bm T_x)_{-D,-D}
\bm B_{-D\cdot}\bm\alpha
\right\}.
\end{aligned}
\end{equation*}

\paragraph{Iterative estimation.}

The source weights and the optimal factors used above are unknown in practice. We estimate $(\bm\omega,\bm\gamma,\bm\alpha)$ using an approximate coordinate-wise algorithm. Given $(\bm\omega^{(t)},\bm\gamma^{(t)},\bm\alpha^{(t)})$ at iteration $t\geq1$, compute $\bm W^{(t)}$, $\bm\Gamma^{(t)}$, and $\bm\Omega^{(t)}$ from \eqref{eq:variational-matrices} by replacing the optimal probabilities with $(\bm\omega^{(t)},\bm\gamma^{(t)})$, and define
\begin{equation}
\bm T^{(t)}
=\mathbb E_{q^{(t)}}(\bm T)
=\bm\Gamma^{(t)}\bm T_x
+\tau_0(\bm I_p-\bm\Gamma^{(t)}).
\label{eq:variational-precision}
\end{equation}
where $q^{(t)}\in\mathcal Q$ has its parameters $(\bm\omega^{(t)},\bm\gamma^{(t)})$.
Update the inclusion probabilities by
\begin{subequations}
\label{eq:probability-iterations-main}
\begin{equation}
\begin{aligned}
\operatorname{logit}(\omega_j^{(t+1)})
={}&\operatorname{logit}(\rho_z)
-\frac12\log
\frac{\left.|\bm T^{(t)}+\bm X^\top\bm X\odot\bm\Omega^{(t)}|\right|_{\omega_j^{(t)}=1}}
     {\left.|\bm T^{(t)}+\bm X^\top\bm X\odot\bm\Omega^{(t)}|\right|_{\omega_j^{(t)}=0}}
\\
&+\left(A_y+\frac{n}{2}\right)
\log\frac{B_y+\frac12\left.\widetilde R(\bm\omega^{(t)},\bm\gamma^{(t)},\bm\alpha^{(t)})\right|_{\omega_j^{(t)}=0}}
{B_y+\frac12\left.\widetilde R(\bm\omega^{(t)},\bm\gamma^{(t)},\bm\alpha^{(t)})\right|_{\omega_j^{(t)}=1}}.
\end{aligned}
\label{eq:omega-update-main}
\end{equation}
With $(\bm\omega^{(t+1)},\bm\gamma^{(t)},\bm\alpha^{(t)})$,
we then update $\bm\gamma^{(t+1)}$ by
\begin{equation}
\begin{aligned}
\operatorname{logit}(\gamma_j^{(t+1)})
={}&\operatorname{logit}(\rho_c)
+\frac12\log
\frac{(\bm T_x)_{jj}}
     {\tau_0}
-\frac12\log
\frac{\left.|\bm T^{(t)}+\bm X^\top\bm X\odot\bm\Omega^{(t+1)}|\right|_{\gamma_j^{(t)}=1}}
     {\left.|\bm T^{(t)}+\bm X^\top\bm X\odot\bm\Omega^{(t+1)}|\right|_{\gamma_j^{(t)}=0}}
\\
&+\left(A_y+\frac{n}{2}\right)
\log\frac{B_y+\frac12\left.\widetilde R(\bm\omega^{(t+1)},\bm\gamma^{(t)},\bm\alpha^{(t)})\right|_{\gamma_j^{(t)}=0}}
{B_y+\frac12\left.\widetilde R(\bm\omega^{(t+1)},\bm\gamma^{(t)},\bm\alpha^{(t)})\right|_{\gamma_j^{(t)}=1}}.
\end{aligned}
\label{eq:gamma-update-main}
\end{equation}
\end{subequations}

For the update of $\bm\alpha$, we follow the variational EM framework
\citep{blei2003latent}.
The part of the ELBO depending on $\bm\alpha$ is
\[
\mathcal L(q^{(t+1)},\bm\alpha)
=
\text{constant}
-\left(A_y+\frac{n}{2}\right)
\mathbb E_{q^{(t+1)}}
\left[
\log\left\{
B_y+\frac12R(\bm z,\bm c,\bm\alpha)
\right\}
\right].
\]
Using the preceding residual approximation gives
\[
\mathbb E_{q^{(t+1)}}
\left[
\log\left\{
B_y+\frac12R(\bm z,\bm c,\bm\alpha)
\right\}
\right]
\approx
\log\left\{
B_y+\frac12\widetilde R(\bm\omega^{(t+1)},\bm\gamma^{(t+1)},\bm\alpha)
\right\}.
\]
Since the logarithm is increasing, maximizing the approximate ELBO is equivalent to minimizing 
$\widetilde R(\bm\omega^{(t+1)},\bm\gamma^{(t+1)},\bm\alpha)$. Expanding the quadratic yields
\begin{equation}
\bm\alpha^{(t+1)}
\in
\operatorname*{arg\,min}_{\bm\alpha\in\Delta_K}
\left\{
\frac12\bm\alpha^\top\bm H_t\bm\alpha
-\bm g_t^\top\bm\alpha
\right\},
\label{eq:alpha-main-qp}
\end{equation}
with
\begin{align}
\bm H_t
&=\bm B^\top\bm\Gamma^{(t+1)}\bm T_x\bm B
-\bm B^\top\bm\Gamma^{(t+1)}\bm T_x
\left\{\bm T^{(t+1)}
+\bm X^\top\bm X\odot\bm\Omega^{(t+1)}\right\}^{-1}
\bm\Gamma^{(t+1)}\bm T_x\bm B,
\label{eq:H-main}
\\*
\bm g_t
&=\bm B^\top
\bm\Gamma^{(t+1)}\bm T_x
\left\{\bm T^{(t+1)}
+\bm X^\top\bm X\odot\bm\Omega^{(t+1)}\right\}^{-1}
\bm W^{(t+1)}\bm X^\top\bm y.
\label{eq:g-main}
\end{align}
Because $
\bm0\preceq
\bm\Gamma^{(t+1)}\bm T_x
\preceq\bm T^{(t+1)}
+\bm X^\top\bm X\odot\bm\Omega^{(t+1)},$
we have $\bm H_t\succeq\bm0$. Then \eqref{eq:alpha-main-qp} is a convex quadratic program.
The source-weight derivation is provided in Section~A.3 of the Supplementary Material.
\paragraph{Initialization.}
We start without source guidance so that arbitrary initial weights do not affect the first inclusion
calculation. Set $\bm\omega^{(0)}=\bm1$, $\bm\gamma^{(0)}=\bm0$, and choose any $\bm\alpha^{(0)}\in\Delta_K$.
The target-only update in \eqref{eq:omega-update-main} at $t=0$ gives $\bm\omega^{(1)}$ independently of 
$\bm\alpha^{(0)}$. Although $\bm\beta$ is integrated out of the variational objective, we use these 
inclusion probabilities to construct an auxiliary target-only coefficient vector $\tilde{\bm\beta}^{(1)}$,
solely to initialize $\bm\alpha^{(1)}$ before updating
$\bm\gamma^{(1)}$:
\begin{equation}
\tilde{\bm\beta}^{(1)}
=
\left\{
\tau_0\bm I_p
+\bm X^\top\bm X\odot\bm\Omega^{(1)}
\right\}^{-1}
\bm W^{(1)}\bm X^\top\bm y.
\label{eq:target-center-main}
\end{equation}
We obtain the initial weights by
matching the source center to $\tilde{\bm\beta}^{(1)}$ through
\begin{equation}
\bm\alpha^{(1)}
\in
\operatorname*{arg\,min}_{\bm\alpha\in\Delta_K}
\sum_{j=1}^p
\omega_j^{(1)}
\ell_{\psi_n}
\left\{
\tilde\beta_j^{(1)}-(\bm B\bm\alpha)_j
\right\},
\label{eq:huber-pre-main}
\end{equation}
where $\psi_n>0$ and

\begin{equation*}
\ell_\psi(u)
=
\begin{cases}
u^2/2, & |u|\leq\psi,\\[2pt]
\psi|u|-\psi^2/2, & |u|>\psi.
\end{cases}
\end{equation*}
Evaluate \eqref{eq:gamma-update-main} at $(\bm\omega^{(1)},\bm\gamma^{(0)},\bm\alpha^{(1)})$
and apply the inverse logit to obtain $\bm\gamma^{(1)}$, completing initialization. 
Subsequent cycles follow \eqref{eq:probability-iterations-main} and \eqref{eq:alpha-main-qp}.
The Huber quadratic-program representation is given in Section~A.3 of the Supplementary Material.
\paragraph{Estimation and variable selection.}
At termination, $t$ is the number of completed cycles. We apply the same matrix substitutions to the 
conditional posterior mean derived in Section~A.1 of the Supplementary Material, and weight
the resulting coefficients by their inclusion probabilities. This gives
\begin{equation}
\begin{aligned}
 \hat{\bm\beta}^{(t)}
 ={}&\bm W^{(t)}
 \left\{\bm T^{(t)}
       +\bm X^\top\bm X\odot\bm\Omega^{(t)}\right\}^{-1}\left\{\bm W^{(t)}\bm X^\top\bm y
       +\bm\Gamma^{(t)}\bm T_x\bm B\bm\alpha^{(t)}\right\}.
\end{aligned}
\label{eq:beta-main-estimator}
\end{equation}
The leading factor $\bm W^{(t)}$ accounts for variable inclusion. We report
\begin{equation}
\begin{gathered}
 \hat{\mathcal S}^{(t)}=\{j:\omega_j^{(t)}>1/2\},\qquad
 \hat{\mathcal C}^{(t)}=\{j:\omega_j^{(t)}>1/2, \gamma_j^{(t)}>1/2\}.
\end{gathered}
\label{eq:selection-sets-main}
\end{equation}
Thus $\hat{\mathcal C}^{(t)}$ contains selected target variables for which source information is 
retained. The algorithm stops when the probability vectors and source weights change by less than 
a specified tolerance, or when a maximum number of cycles is reached. Algorithm~\ref{alg:collapsed_vb} 
summarizes the procedure.

\begin{algorithm}[H]
\caption{TRACE: approximated collapsed variational algorithm}
\label{alg:collapsed_vb}
\DontPrintSemicolon
\Input{$\bm X,\bm y,\bm B,n_s$; model hyperparameters;
Huber threshold $\psi_n$; tolerance $\varepsilon$; maximum cycles $L$.}
\Output{$\hat{\bm\beta}^{(t)}$, $\bm\omega^{(t)}$, $\bm\gamma^{(t)}$,
$\bm\alpha^{(t)}$ and selected variable and transfer sets.}
Set $\bm\omega^{(0)}=\bm1$, $\bm\gamma^{(0)}=\bm0$ and
$\bm\alpha^{(0)}\in\Delta_K$; compute $\bm T_x$ from
\eqref{eq:prior-precisions-main}\;
Compute $\bm\omega^{(1)}$ by \eqref{eq:omega-update-main} at $t=0$
and update its matrix expectations by \eqref{eq:variational-matrices}\;
Compute $\tilde{\bm\beta}^{(1)}$ by \eqref{eq:target-center-main}
and $\bm\alpha^{(1)}$ by \eqref{eq:huber-pre-main}\;
Compute $\bm\gamma^{(1)}$ by \eqref{eq:gamma-update-main} using
$(\bm\omega^{(1)},\bm\gamma^{(0)},\bm\alpha^{(1)})$; set $t=1$\;
\While{$t<L$}{
 Update $\bm\omega^{(t+1)}$ by \eqref{eq:omega-update-main}\;
 Update $\bm\gamma^{(t+1)}$ by \eqref{eq:gamma-update-main}
 using the new inclusion probabilities\;
 Compute the required matrices by \eqref{eq:variational-matrices}
 and \eqref{eq:variational-precision}, then update
 $\bm\alpha^{(t+1)}$ by \eqref{eq:alpha-main-qp}--\eqref{eq:g-main}\;
 Set $t\gets t+1$; stop if the largest coordinatewise change in
 $(\bm\omega,\bm\gamma,\bm\alpha)$ is below $\varepsilon$\;
}
Compute \eqref{eq:beta-main-estimator} and \eqref{eq:selection-sets-main}\;
\end{algorithm}

\section{Theoretical Properties}
\label{sec:theory}
We study the theoretical properties of TRACE in fixed and growing dimensions. The fixed-dimensional 
results concern the mathematical updates in Algorithm~\ref{alg:collapsed_vb} over any prespecified
finite number of complete cycles. The high-dimensional results concern the exact conditional collapsed 
posterior and global maximizers of the collapsed ELBO.

In both settings, the target data are generated from the true model
\[
\bm y=\bm X\bm\beta_0+\bm\varepsilon,
\qquad
\bm\varepsilon\mid\bm X
\sim\mathcal N_n(\bm0,\sigma_0^2\bm I_n),
\qquad
0<\sigma_0^2<\infty,
\]
where $\bm\beta_0$ is the true target coefficient vector. The population source coefficient matrix 
and weight vector are denoted by $\bm B_0$ and $\bm\alpha_0\in\Delta_K$, respectively. Define the 
true target support and its transferable subset by
\[
\mathcal S_0=\{j\in[p]:\beta_{0j}\ne0\},
\qquad
\mathcal C_0=
\{j\in\mathcal S_0:(\bm B_0\bm\alpha_0)_j=\beta_{0j}\}.
\]
Write $s=|\mathcal S_0|$. The random selected support is $\mathcal S=\{j\in[p]:z_j=1\}.$ For $u\subseteq[p]$, define the transfer set within $u$ by $\mathcal C_u=\{j\in u:c_j=1\}.$ Recovery of the transfer structure 
is assessed by whether $\mathcal C_{\mathcal S_0}=\mathcal C_0$, since the likelihood does not identify 
$c_j$ when $z_j=0$. The orthogonal projector onto the column space of $\bm X_u$ is
\[
\bm P_u
=\bm X_u(\bm X_u^\top\bm X_u)^\dagger\bm X_u^\top,
\qquad
\bm P_{\varnothing}=\bm0,
\]
where $^\dagger$ denotes the Moore--Penrose inverse.

The following conventions apply to the asymptotic statements below. We use $|D|$ for set cardinality 
and $a\vee b=\max(a,b)$. For positive sequences, $a_n\lesssim b_n$ means $a_n\leq Cb_n$ for some constant 
$C>0$ and all sufficiently large $n$. The relation $a_n\asymp b_n$ means $cb_n\leq a_n\leq Cb_n$ 
for some constants $0<c<C<\infty$ and all sufficiently large $n$. Constants are independent of $n$ 
unless stated otherwise. In sampling statements, $\Prob$, $\mathbb E$, $O_p$, $o_p$, and convergence in probability 
refer to the joint true sampling law of the target data and source summaries.

\subsection{Fixed-dimensional consistency}
\label{subsec:fixed-theory-main}
We consider fixed $p\geq2$, $K$, and population parameters, with $n,n_s\to\infty$. Define
\[
\begin{aligned}
M_B
&:=\max_{j\in[p]}\|(\bm B_0)_{j\cdot}\|_2<\infty,\\
C_b
&:=\min_{j\in\mathcal S_0\setminus\mathcal C_0}
\bigl|(\bm B_0\bm\alpha_0)_j-\beta_{0j}\bigr|>0,
\quad\text{if }\mathcal S_0\setminus\mathcal C_0\ne\varnothing.
\end{aligned}
\]
\paragraph{Assumption A.}
\begin{enumerate}[itemsep=0.75ex,parsep=0pt]
\item[(A1)] The fixed population second-moment matrix $\bm\Sigma_x=\mathbb E(n^{-1}\bm X^\top\bm X)$ 
satisfies $\bm\Sigma_x\succ\bm0$ and $\bm X^\top\bm X=n\bm\Sigma_x+o_p(n)$.

\item[(A2)] The average source sample size satisfies $n_s\asymp n$.

\item[(A3)] The set $\mathcal C_0$ is nonempty, and the source summaries satisfy
$\max_{j\in[p]}\|\bm B_{j\cdot}-(\bm B_0)_{j\cdot}\|_2=O_p(n_s^{-1/2})$. Moreover,
\begin{align*}
\inf_{\substack{\bm\alpha\in\Delta_K\\
                \bm\alpha\neq\bm\alpha_0}}
&\frac{\sum_{j\in\mathcal C_0}
\{(\bm B_0)_{j\cdot}(\bm\alpha-\bm\alpha_0)\}^2}
{|\mathcal C_0|\,\|\bm\alpha-\bm\alpha_0\|_2^2}
\geq\kappa_B^2>0,\\
&\frac{|\mathcal S_0\setminus\mathcal C_0|}{|\mathcal S_0|}
\leq\bar\pi<
\frac{\kappa_B^2}{\kappa_B^2+M_B^2}.
\end{align*}
\item[(A4)]
The hyperparameters satisfy
\[
\rho_z\lesssim n^{-1},\qquad -\operatorname{logit}(\rho_z)=o(n),\qquad
\rho_c\asymp n_s^{-1/4},\qquad \psi_n\asymp n_s^{-1/2}.
\]

The hyperparameters $A_y$, $B_y$, and $\tau_0$ are positive constants.
\end{enumerate}

Condition~(A1) ensures a nonsingular population design. Condition~(A2) makes the target and source 
sample sizes comparable. Condition~(A3) controls source-summary error and provides identification 
of the source weights. The contamination bound ensures that compatible coordinates dominate incompatible 
ones in the Huber initialization. Condition~(A4) specifies the prior scales governing target sparsity
and the size of the transfer set, together with the Huber threshold $\psi_n$ for robust initialization 
of the source weights.

\begin{theorem}
\label{thm:fixed-main}
Under Assumption~A, initialize $\bm\omega^{(0)}=\bm1$ and $\bm\gamma^{(0)}=\bm0$. For every prespecified 
fixed integer $t\geq1$, there exist constants $c_\omega>0$ and $c_\gamma>0$ such that the following 
bounds hold for each nonempty coordinate class:
\begin{align*}
\omega_j^{(t)}
&=\begin{cases}
    1-O_p\{\exp(-c_\omega n)\}, \quad & j\in \mathcal{S}_0, \\
   O_p(n^{-1}), \quad &j\notin\mathcal S_0,\\
\end{cases}\\
\gamma_j^{(t)}
&=\begin{cases}
 1- O_p(n_s^{-1/4}), \quad & j\in\mathcal C_0,\\
 O_p\{\exp(-c_\gamma n)\}, \quad &  j\in \mathcal S_0\setminus\mathcal C_0,\\
 O_p(n_s^{-1/4}) ,\quad &  j\notin\mathcal S_0
 \end{cases}\\
\|\bm\alpha^{(t)}&-\bm\alpha_0\|_2
=O_p(n_s^{-1/2}).
\end{align*}
\end{theorem}

\begin{corollary}

\label{cor:fixed-selection-main}

Under the conditions of Theorem~\ref{thm:fixed-main}, for every prespecified fixed integer $t\geq1$, 
the estimates obtained by thresholding at $1/2$ satisfy
\[
\Prob\{\hat{\mathcal S}^{(t)}=\mathcal S_0,\,
       \hat{\mathcal C}^{(t)}=\mathcal C_0\}\longrightarrow1.
\]
\end{corollary}

Under Assumption~A, Theorem~\ref{thm:fixed-main} establishes selection consistency for the target variables and consistency of transfer decisions on active coordinates at each fixed iteration. The rate bounds are obtained by induction over the update cycles of Algorithm~\ref{alg:collapsed_vb}. Proofs of the theorem and Corollary~\ref{cor:fixed-selection-main} are given in Section~B of the Supplementary Material.

Next, we examine how compatible source information affects the inclusion probabilities for active variables. Under the conditions below, the leading asymptotic term of the inclusion log-odds is no smaller under TRACE than under target-only ACVB.
\begin{corollary}
\label{cor:fixed-rate-main}
Under the conditions of Theorem~\ref{thm:fixed-main}, suppose additionally that $n_s/n\to\eta\in(0,\infty)$. For every $j\in\mathcal S_0$ and every prespecified fixed integer $t\geq2$, $\omega_j^{\mathrm{TRACE},(t)}$ converges to one at least as fast as $\omega_j^{\mathrm{ACVB},(t)}$ on the exponential scale.
\end{corollary}
The proof is given in Section~B.4 of the Supplementary Material.

\subsection{ High dimensional consistency}
\label{subsec:high-theory-main}
Now allow $p=p_n\to\infty$. Let $(q^\ast,\bm\alpha^\ast)$ denote any global maximizer of the exact collapsed ELBO in \eqref{eq:collapsed-elbo-main} over $\mathcal Q\times\Delta_K$, with $\mathcal Q$ defined in \eqref{eq:mean-field-main}. For fixed $\bm\alpha\in\Delta_K$, write $\Prob_{\bm\alpha}(\,\cdot\mid\bm y)$ for the exact collapsed posterior of $(\bm z,\bm c)$ conditional on that source weight.

\begin{proposition}
\label{prop:global-main}

For every fixed $n$ and $p$, the exact collapsed ELBO is continuous on $[0,1]^{2p}\times\Delta_K$. 
Hence
\[
\arg
\max_{\substack{(\bm\omega,\bm\gamma)\in[0,1]^{2p}\\\bm\alpha\in\Delta_K}}
\mathbb E_q\log\frac{p(\bm y\mid\bm z,\bm c,\bm\alpha)\Prob(\bm z)\Prob(\bm c)}{q(\bm z,\bm c)}
\neq\varnothing.
\]
\end{proposition}

Proposition~\ref{prop:global-main} ensures that, for each sample size and dimension, the exact collapsed ELBO attains its maximum over the feasible set. Thus, the global variational EM optimizer is well defined.

\paragraph{Assumption B.}
\begin{enumerate}[itemsep=0.75ex,parsep=0pt]

\item[(B1)]
As $n\to\infty$, $p=p_n\to\infty$, $n_s\asymp n$, with $K<\infty$ fixed, we have $s^2\log p=o(n_s)$.
\item[(B2)]
For some $0<C_1<C_2<\infty$ and $M_\beta<\infty$, $C_1\leq\sigma_0^2\leq C_2$ and
$\max_{j\in\mathcal S_0}|\beta_{0j}|\leq M_\beta$. For sufficiently large $C_\beta>0$,
\[\min_{j\in\mathcal S_0}\beta_{0j}^2\geq C_\beta\frac{\log(p\vee\sqrt n)}{n}.\]
\item[(B3)]
For some $0<C_3<C_4<\infty$, $C_3\leq\|\bm X_j\|_2^2/n\leq C_4$ for every $j\in[p]$ and $\lambda_{\max}(\bm X_{\mathcal S_0}^\top \bm X_{\mathcal S_0}/n)\leq C_4$. For sufficiently large $K_0>1$ and some $C_5>0$,
\[
\inf_{\substack{u\subseteq\mathcal S_0^c\\|u|\leq K_0s}}
\lambda_{\min}\left\{n^{-1}\bm X_{\mathcal S_0}^\top
(\bm I-\bm P_u)\bm X_{\mathcal S_0}\right\}\geq C_5.
\]
\item[(B4)]
The source coefficients satisfy
$|\mathcal C_0|\geq1$ and $\max_{1\leq j\leq p}\|(\bm B_0)_{j\cdot}\|_2\leq M_B$
for some $M_B<\infty$. If $\mathcal S_0\setminus\mathcal C_0\ne\varnothing$,
$\min_{j\in\mathcal S_0\setminus\mathcal C_0} |(\bm B_0\bm\alpha_0)_j-\beta_{0j}|\geq C_b>0$. Moreover, for every $\mathcal H\subseteq\mathcal S_0$ with $|\mathcal H|=|\mathcal C_0|$,
\[
\inf_{\substack{\bm\alpha\in\Delta_K\\\bm\alpha\neq\bm\alpha_0}}
\frac{|\mathcal H|^{-1}\sum_{j\in\mathcal H}\{\beta_{0j}-(\bm B_0)_{j\cdot}\bm\alpha\}^2}{\|\bm\alpha-\bm\alpha_0\|_2^2}
\geq\kappa_B^2>0.
\]
The source summaries $\bm B$, based on samples independent of $(\bm y,\bm X)$, satisfy
\[
\begin{aligned}
\max_{1\leq j\leq p}\|\bm B_{j\cdot}-(\bm B_0)_{j\cdot}\|_2
&=O_p\!\left(\sqrt{\frac{\log p}{n_s}}\right),\\
\|(\bm B_{\mathcal C_0\cdot}-(\bm B_0)_{\mathcal C_0\cdot})\bm\alpha_0\|_2^2
&=O_p\!\left(\frac{\log p}{n_s}\right).
\end{aligned}
\]
\item[(B5)]
The priors on target inclusion and information transfer satisfy $\rho_z=\{p(p^{3/2}\vee\sqrt n)\}^{-1}$ and $0<C_6\leq\rho_c\leq C_7<1$. The hyperparameters $A_y$, $B_y$, and $\tau_0$ are positive constants.
\end{enumerate}

Condition~(B1) controls model complexity relative to sample sizes. Condition~(B2) bounds the noise variance and requires sufficient signal strength for consistent variable selection. Condition~(B3) controls predictor scales and preserves signal information after adjustment for inactive predictors. Condition~(B4) controls source--target discrepancy on $\mathcal C_0$ as in Assumption~1 (iii) of \citet{zhang2024concert}. It also requires a positive gap on $\mathcal S_0\setminus\mathcal C_0$ and ensures identification of the source weights $\bm\alpha$. Condition~(B5) specifies the prior probabilities for variable inclusion and information transfer.
\begin{theorem}
\label{thm:high-alpha-main}
Under Assumption~B, for the source-weight component $\bm\alpha^\ast$ of any global variational EM maximizer,
\[
\|\bm\alpha^\ast-\bm\alpha_0\|_2
=O_p\left(\sqrt{\frac{\log p}{|\mathcal C_0|n_s}}\right).
\]
\end{theorem}
Theorem~\ref{thm:high-alpha-main} establishes consistency of the source-weight estimator. The convergence rate improves with the source sample size and the number of transferable active coordinates. The proof is given in Section~C.5 of the Supplementary Material.


We next use the above source-weight bound to study variable selection and transfer decisions at $\bm\alpha^\ast$. The analysis starts with the exact collapsed posterior and uses the posterior-ratio approach of \citet{narisetty2014variable}.

\begin{theorem}
\label{thm:high-posterior-main}

Under Assumption~B,
\[
\Prob_{\bm\alpha^\ast}
\left(
\mathcal S=\mathcal S_0,
\mathcal C_{\mathcal S_0}=\mathcal C_0
\mid\bm y
\right)
\overset{P}{\longrightarrow}1.
\]
\end{theorem}

Theorem~\ref{thm:high-posterior-main} establishes joint posterior consistency for the target signal set and transfer states on the active coordinates at $\bm\alpha^\ast$. The proof is given in the corresponding subsection of Section~C of the Supplementary Material. We next consider the global mean-field approximation.

\begin{theorem}
\label{thm:high-vb-main}
Under Assumption~B, let $(q^\ast,\bm\alpha^\ast)$ be any global maximizer in Proposition~\ref{prop:global-main}. Then
\[
q^\ast
\left(
\mathcal S=\mathcal S_0,
\mathcal C_{\mathcal S_0}=\mathcal C_0
\right)
\overset{P}{\longrightarrow}1.
\]
Therefore, for each nonempty coordinate class,
\[
\max_{j\in\mathcal S_0}(1-\omega_j^\ast)\overset{P}{\longrightarrow}0,
\qquad
\max_{j\notin\mathcal S_0}\omega_j^\ast\overset{P}{\longrightarrow}0,
\]
and
\[
\max_{j\in\mathcal C_0}(1-\gamma_j^\ast)\overset{P}{\longrightarrow}0,
\qquad
\max_{j\in\mathcal S_0\setminus\mathcal C_0}\gamma_j^\ast\overset{P}{\longrightarrow}0.
\]
\end{theorem}

Theorem~\ref{thm:high-vb-main} shows that the global mean-field
approximation retains the joint selection consistency of the exact
posterior. Beyond selection consistency, we examine how compatible source information affects the evidence for including a true active variable. The following corollary compares the determinant contributions with and without transfer.
\begin{corollary}
\label{cor:high-det-gain-main}
Under Assumption~B, fix $j\in\mathcal C_0$ and suppose that the other true active variables are included with their transfer indicators fixed at their true values. Using source information for $j$ then increases the determinant contribution to its inclusion evidence by a factor of order $\sqrt{n/\log p}$ compared with setting $c_j=0$.
\end{corollary}

The proof of Theorem~\ref{thm:high-vb-main} and Corollary~\ref{cor:high-det-gain-main} are provided in Section~C of the Supplementary Material. The numerical studies below complement the theoretical analysis by evaluating the practical performance of the coordinate algorithm in Section~\ref{subsec:approx-cvb-main}, using a numerical stopping rule.

\section{Simulation}
\label{sec:simulation}
We evaluate variable selection and source-weight estimation under four patterns of source heterogeneity. The comparisons assess gains over target-only analysis, sensitivity to incompatible source information, and the effect of estimating the source weights. We compare TRACE-EM, implemented as in Section~\ref{subsec:approx-cvb-main}, with Trans-Lasso \citep{li2022transfer}, TRADER \citep{lai2026bayesian}, and CONCERT \citep{zhang2024concert}. ACVB \citep{you2023ollapsed} provides the target-only benchmark obtained by disabling information transfer in TRACE. TRACE and TRADER receive the same fitted source coefficient summaries. Trans-Lasso and CONCERT use the observations from which those summaries are fitted. To assess source-weight estimation, we also include Oracle TRACE, which fixes $\bm\alpha=\bm\alpha_0$ and retains the other TRACE updates. The code and data are available at \url{https://github.com/Sivanfia/TRACE}.

Setting~1 considers contamination shared across sources. Setting~2 uses the TRADER benchmarks in \citet{lai2026bayesian} which varies the number of informative sources. Setting~3 allows the locally compatible source to change across variables which favors the coordinate-specific source selection used by CONCERT \citep{zhang2024concert}. Setting~4 considers complementary sources whose individual deviations cancel under an appropriate weighted combination.

TRACE-EM, Oracle TRACE, ACVB, and CONCERT select variables using a posterior inclusion threshold of $0.5$. TRADER uses a posterior same-sign probability of at least $95\%$. Trans-Lasso selection uses an absolute-value threshold of $10^{-4}$ on the target-only refitted coefficients. The primary criterion is $F_1$ for variable selection,
\[
\mathrm{F_1}=\frac{2|\hat{\mathcal S}\cap\mathcal S_0|}
{|\hat{\mathcal S}|+|\mathcal S_0|},
\]
where $\hat{\mathcal S}$ is the selected variable set for each method.

Every parameter combination has 50 replications. Figures report means with one Monte Carlo standard error. Additional results and diagnostics for the four settings are given in Sections~D.1--D.4 of the Supplementary Material.

\subsection*{Setting 1: shared active-coordinate contamination}
Target observations follow $y_i=\bm x_i^\top\bm\beta_0+\varepsilon_i$, with $\bm x_i\sim N_p(\bm0,\bm\Sigma_x)$, $(\bm\Sigma_x)_{j\ell}=\rho^{|j-\ell|}$, and independent $\varepsilon_i\sim N(0,\sigma^2)$. Set $n=300$, $p=600$, $K=3$, and $\beta_{0j}=(-1)^{j+1}\mathbb I(j\leq s)$. The grid uses $s\in\{10,15,30\}$, $\rho\in\{0.3,0.6\}$, and $\mathrm{SNR}\in\{3,4\}$, with $\sigma^2=\bm\beta_0^\top\bm\Sigma_x\bm\beta_0/\mathrm{SNR}$.
For $\bm\alpha_0=(0.5,0.3,0.2)^\top$, define $m=\operatorname{round}(qs)$, $q\in\{0,0.2,0.4,0.6\}$, and
\[
(\beta_j^{(1)},\beta_j^{(2)},\beta_j^{(3)})=
\begin{cases}
(0,-1,-2)\beta_{0j}, & 1\leq j\leq m,\\
(1.25,0.85,0.60)\beta_{0j}, & m<j\leq s,\quad j-m\text{ odd},\\
(0.80,1.30,1.05)\beta_{0j}, & m<j\leq s,\quad j-m\text{ even},\\
(0,0,0), & j>s.
\end{cases}
\]
Thus $\mathcal C_0=\{m+1,\ldots,s\}$, and the weighted source center is $-0.7\beta_{0j}$ on the contaminated coordinates. Each source has $n_s=500$ observations generated with the target
covariate covariance and error variance. TRACE and TRADER receive the same spike-and-slab-based coefficient summaries, fitted using \texttt{spikeslab} \citep{ishwaran2022spikeslab} and expressed on the original predictor scale. CONCERT and Trans-Lasso use the corresponding source observations.

\begin{figure}[H]
\centering
\includegraphics[width=\textwidth]{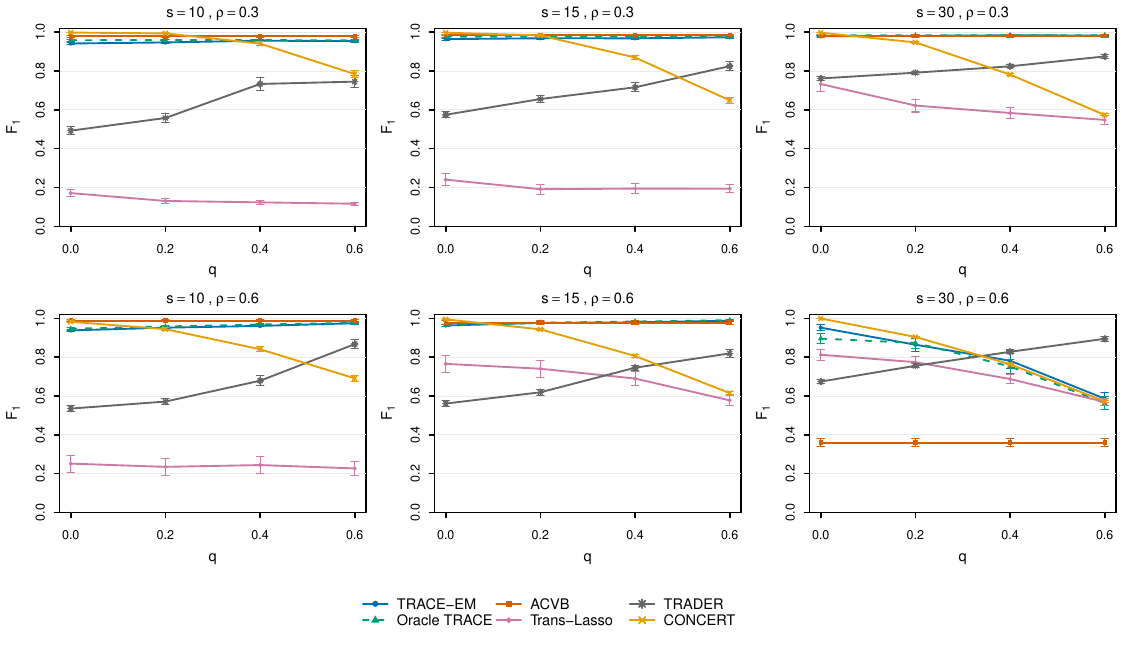}
\caption{Mean $F_1$ ($\pm1$ Monte Carlo standard error) in Setting~1 for $s\in\{10,15,30\}$,
$\rho\in\{0.3,0.6\}$, and contamination fraction $q\in\{0,0.2,0.4,0.6\}$. All panels use $n=300$, $p=600$, $\mathrm{SNR}=4$, and 500 observations per source.}
\label{fig:simulation-setting1}
\end{figure}

Figure~\ref{fig:simulation-setting1} shows that TRACE-EM maintains high mean $F_1$ across contamination levels at $s=\{10,15\}$, with performance close to ACVB. At $s=30$ and $\rho=0.6$, TRACE-EM improves on ACVB throughout the displayed contamination range, despite greater
sensitivity to contamination. TRACE has higher mean $F_1$ than Trans-Lasso in all displayed configurations and than TRADER in most. Its advantage over CONCERT is generally more apparent at higher contamination levels. TRACE-EM and Oracle TRACE show similar selection trends. 
The $\mathrm{SNR}=3$ results are reported in Supplementary Section~D.1.

\subsection*{Setting 2: TRADER benchmark designs}
Following Settings~I and~III in \citet{lai2026bayesian}, we use $K=10$ sources and vary the number of informative sources $K_a\in\{1,\ldots,10\}$. Table~\ref{tab:trader-simulation-designs} summarizes the parameters, where $n_k$ is the sample size of each source and $h$ controls source perturbation. The true target model is $\bm y=\bm X\bm\beta_0+\bm\varepsilon$, $\bm\varepsilon\mid\bm X\sim\mathcal N_n(\bm0,\bm I_n)$, with $\beta_{0j}=0.5\,\mathbb I(j\in\mathcal S_0)$ and $\mathcal S_0=\{1,\ldots,s\}$. Target covariates are Gaussian with $(\bm\Sigma_x)_{j\ell}=0.5^{|j-\ell|}$.

\begin{table}[htbp]
\centering
\caption{Parameters for the two TRADER benchmark designs in Setting 2.}
\label{tab:trader-simulation-designs}
\begin{tabular}{lrrrrr}
\toprule
Design & $n$ & $p$ & $s$ & $n_k$ & $h$\\
\midrule
TRADER Setting I & 200 & 500 & 5 & 200 & 5\\
TRADER Setting III & 100 & 2000 & 20 & 500 & 10\\
\bottomrule
\end{tabular}
\end{table}

Informative sources $k\leq K_a$ have coefficients $\beta_j^{(k)}=\beta_{0j}+(h/p)\xi_{jk}$,
where $\xi_{jk}$ are independent Rademacher variables. Noninformative sources have coefficients $\beta_j^{(k)}=0.5\,\mathbb I(j\in\mathcal H_k) +(2h/p)\xi_{jk}$, where $\mathcal H_k$ contains $\{s+1,\ldots,2s\}$ and $s$ additional indices sampled from $\{2s+1,\ldots,p\}$. They therefore introduce strong signals on $2s$ target-null coordinates. Source covariates have independent Student $t_4$ entries, with independent $N(0,1)$ errors in the source linear models. The reference weights are $\alpha_{0k}=K_a^{-1}\mathbb I(k\leq K_a)$. TRACE and TRADER receive the same fitted spike-and-slab source coefficient summaries, while CONCERT and Trans-Lasso use individual-level source observations.

\begin{figure}[htbp]
\centering
\includegraphics[width=\textwidth]{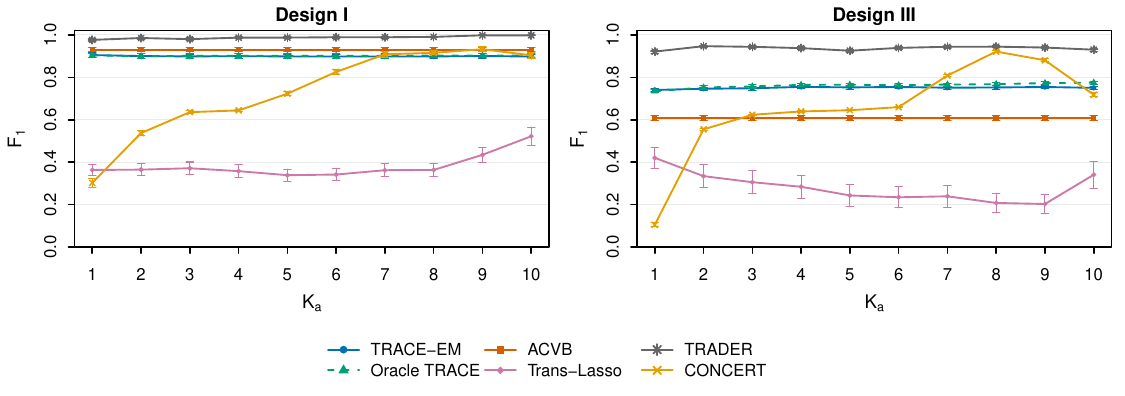}
\caption{Mean $F_1$ ($\pm1$ Monte Carlo standard error) in Setting~2 for $K_a\in\{1,\ldots,10\}$ informative sources in Designs~I and~III. Both designs use $K=10$ sources; the remaining parameters are given in Table~\ref{tab:trader-simulation-designs}.}
\label{fig:simulation-setting2}
\end{figure}

Figure~\ref{fig:simulation-setting2} shows that TRADER achieves the highest mean $F_1$ in both benchmarks, with little variation across $K_a$. TRACE-EM also provides stable and competitive performance. In Design~I, its mean $F_1$ is only slightly lower than that of ACVB; while in Design~III, TRACE improves both precision and recall over ACVB, resulting in higher $F_1$. CONCERT is more sensitive to $K_a$ and has lower mean $F_1$ than TRACE-EM in most configurations. TRACE-EM and Oracle TRACE achieve similar mean $F_1$ in both designs.

\subsection*{Setting 3: coordinate-wise heterogeneous source contamination}
Setting~3 uses the target model, parameter grid, and source-sample construction of Setting~1, 
with $\bm\alpha_0=(1/3,1/3,1/3)^\top$. For $m=\operatorname{round}(qs)$, transferable coordinates $m<j\leq s$ alternate between source multipliers $(1.25,0.85,0.90)$ and $(0.80,1.30,0.90)$, both with weighted mean one. On $1\leq j\leq m$, the multipliers cycle through $(1,-2,-2)$, $(-2,1,-2)$, and $(-2,-2,1)$. To introduce source-only signals, sample disjoint sets $\mathcal U_1,\mathcal U_2,\mathcal U_3\subseteq\mathcal S_0^c$ with total size $\operatorname{round}(m/2)$ and sizes differing by at most one. For each source $k$, set $(\bm\beta^{(k)})_{\mathcal U_k}=(-1,1,-1,\ldots)^\top$, $(\bm\beta^{(k)})_{\mathcal S_0^c\setminus\mathcal U_k}=\bm0$, with entries ordered by predictor index. Each added signal occurs in one source and is absent 
from the target.

\begin{figure}[htbp]
\centering
\includegraphics[width=\textwidth]{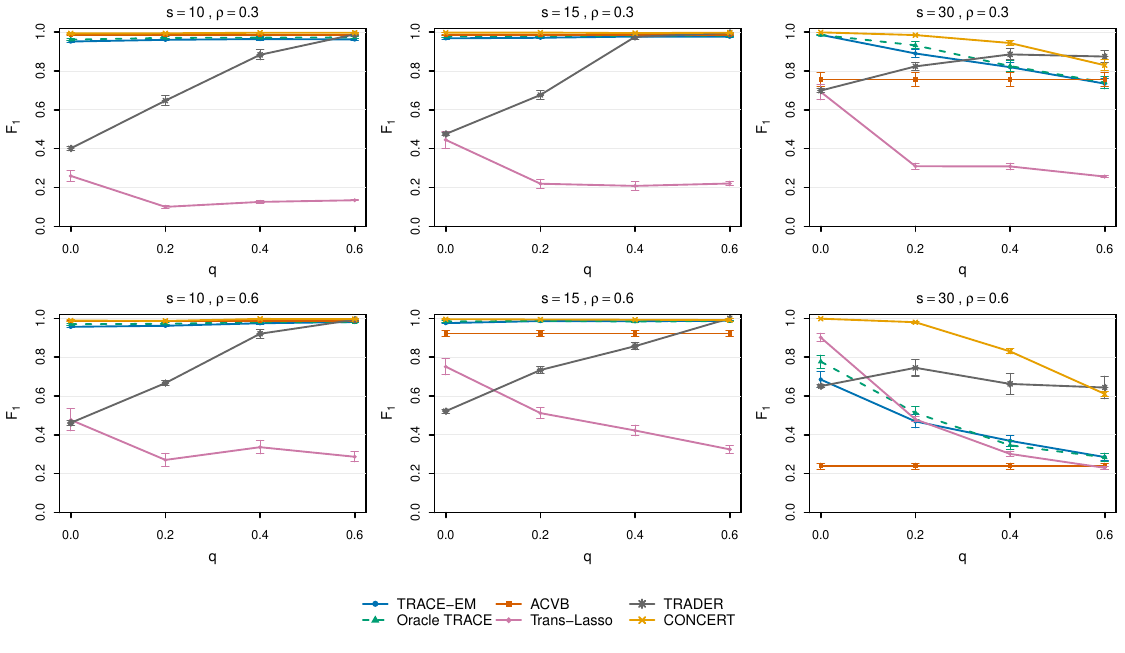}
\caption{Mean $F_1$ ($\pm1$ Monte Carlo standard error) in Setting~3 for $s\in\{10,15,30\}$, $\rho\in\{0.3,0.6\}$, and contamination fraction $q\in\{0,0.2,0.4,0.6\}$. All panels use $n=300$, $p=600$, $\mathrm{SNR}=3$, and 500 observations per source.}
\label{fig:simulation-setting3}
\end{figure}

Figure~\ref{fig:simulation-setting3} shows that TRACE-EM maintains high mean $F_1$ and performs close to CONCERT in most configurations, with CONCERT generally achieving the highest mean $F_1$. TRACE-EM becomes more sensitive to contamination at $s=30$ and $\rho=0.6$. TRADER becomes more competitive as contamination increases, narrowing the gap with the leading methods in several configurations. Relative to ACVB, TRACE-EM generally achieves comparable or higher mean $F_1$. The $\mathrm{SNR}=4$ results are reported in Supplementary Section~D.3.

\subsection*{Setting 4: complementary sources with shared contamination}
The target follows the Gaussian design and has the alternating unit-magnitude coefficients of Setting~1, with $n=300$, $p=600$, $K=3$, $\rho=0.3$, $s\in\{10,15,30\}$, and independent $\varepsilon_i\sim N(0,9)$. Each source has 500 observations. For $\bm\alpha_0=(0.5,0.3,0.2)^\top$, let $m=\operatorname{round}(qs)$, $q\in\{0,0.1,\ldots,0.6\}$. The source multipliers on $m<j\leq s$ cycle by predictor index through $(3,-1,-1)$, $(-0.2,3.8,-0.2)$ and $(0.2,0.2,4.2)$, each giving weighted coefficient $\beta_{0j}$. All sources have coefficient $-\beta_{0j}$ on $1\leq j\leq m$. For a randomly selected set $\mathcal U\subseteq\mathcal S_0^c$ with $|\mathcal U|=s$, set $(\bm\beta^{(k)})_{\mathcal U}=(-1,1,-1,\ldots)^\top$, $k=1,2,3$, with entries ordered by predictor index and all other null coefficients equal to zero. Source observations and coefficient summaries are generated and fitted as in Setting~1.

\begin{figure}[htbp]
\centering
\includegraphics[width=\textwidth]{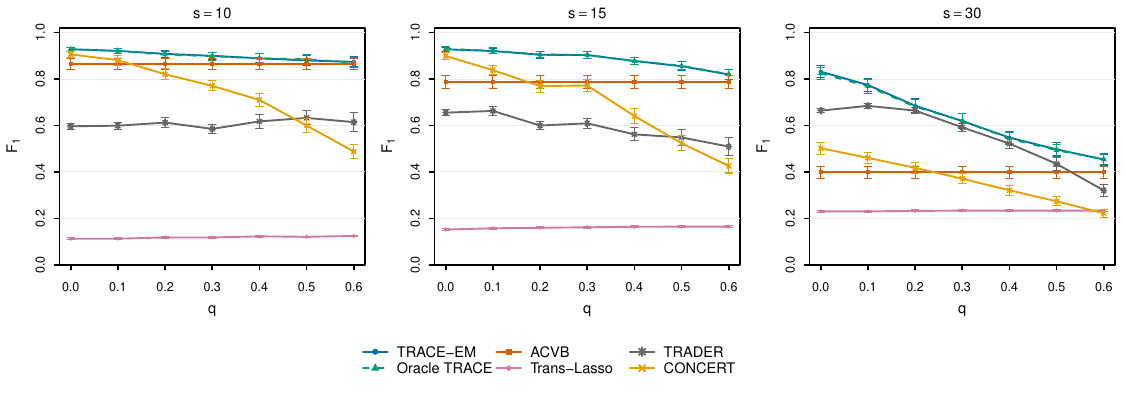}
\caption{Mean $F_1$ ($\pm1$ Monte Carlo standard error) in Setting~4 for $s\in\{10,15,30\}$, $\rho=0.3$, and contamination fraction $q\in\{0,0.1,\ldots,0.6\}$. All panels use $n=300$, $p=600$, noise standard deviation $\sigma=3$, and 500 observations per source.}
\label{fig:simulation-setting4}
\end{figure}

The complementary-source setting provides the clearest gains from TRACE. Figure~\ref{fig:simulation-setting4} shows higher mean $F_1$ than ACVB and all three transfer-learning competitors in all configurations. TRACE-EM closely tracks Oracle TRACE. The largest absolute difference in mean selection $F_1$ is $0.0076$.

\section{Real Data Application}
\label{sec:real-data}
We follow the GTEx v8 SH2D2A expression-prediction example of \citet{zhang2024concert}. The response 
is $\log_2(\mathrm{TPM}+1)$ expression of SH2D2A, with other genes in the same central-nervous-system 
module as predictors. Each of six brain tissues is treated as the target, with the remaining five 
tissues as sources. For each target, we repeat an 80/20 donor-level train--test split 50 times, 
excluding target test donors from all sources. All screening and standardization use training data 
only. TRACE and TRADER receive fitted ridge source summaries and bootstrap-calibrated effective 
source sample sizes, while Trans-Lasso and CONCERT use individual-level source data.
Each selected model is refitted by target-training ridge regression with a cross-validated penalty.
Training donors may overlap between target and source tissues. Table~\ref{tab:gtex-mse} reports test MSE 
on the original $\log_2(\mathrm{TPM}+1)$ scale. TRACE achieves the lowest mean MSE for amygdala.
CONCERT achieves the lowest mean MSE in the remaining tissues. Moreover, Trans-Lasso is also competitive
in hippocampus, nucleus accumbens, and substantia nigra.

\begin{table}[H]
\centering
\caption{Held-out prediction MSE for SH2D2A expression. Entries are mean (standard deviation) over 50 donor-level splits; smaller values are better. All methods use a common target-only post-selection ridge refit.}
\label{tab:gtex-mse}

\setlength{\tabcolsep}{3pt}
\begin{tabular}{lccccc}
\toprule
Target tissue & TRACE & ACVB & TRADER & Trans-Lasso & CONCERT \\
\midrule

Amygdala & \begin{tabular}[c]{@{}c@{}}\textbf{0.0232}\\\textbf{(0.0083)}\end{tabular} & \begin{tabular}[c]{@{}c@{}}0.0238\\(0.0094)\end{tabular} & \begin{tabular}[c]{@{}c@{}}0.0347\\(0.0231)\end{tabular} & \begin{tabular}[c]{@{}c@{}}0.0307\\(0.0099)\end{tabular} & \begin{tabular}[c]{@{}c@{}}0.0258\\(0.0134)\end{tabular}\\

Anterior cingulate cortex & \begin{tabular}[c]{@{}c@{}}0.0415\\(0.0251)\end{tabular} & \begin{tabular}[c]{@{}c@{}}0.0419\\(0.0239)\end{tabular} & \begin{tabular}[c]{@{}c@{}}0.0543\\(0.0272)\end{tabular} & \begin{tabular}[c]{@{}c@{}}0.0404\\(0.0220)\end{tabular} & \begin{tabular}[c]{@{}c@{}}\textbf{0.0350}\\\textbf{(0.0171)}\end{tabular}\\
Cortex & \begin{tabular}[c]{@{}c@{}}0.0420\\(0.0101)\end{tabular} & \begin{tabular}[c]{@{}c@{}}0.0412\\(0.0100)\end{tabular} & \begin{tabular}[c]{@{}c@{}}0.0520\\(0.0163)\end{tabular} & \begin{tabular}[c]{@{}c@{}}0.0401\\(0.0088)\end{tabular} & \begin{tabular}[c]{@{}c@{}}\textbf{0.0389}\\\textbf{(0.0078)}\end{tabular}\\
Hippocampus & \begin{tabular}[c]{@{}c@{}}0.0350\\(0.0087)\end{tabular} & \begin{tabular}[c]{@{}c@{}}0.0363\\(0.0096)\end{tabular} & \begin{tabular}[c]{@{}c@{}}0.0516\\(0.0127)\end{tabular} & \begin{tabular}[c]{@{}c@{}}0.0308\\(0.0108)\end{tabular} & \begin{tabular}[c]{@{}c@{}}\textbf{0.0300}\\\textbf{(0.0093)}\end{tabular}\\
Nucleus accumbens & \begin{tabular}[c]{@{}c@{}}0.0417\\(0.0152)\end{tabular} & \begin{tabular}[c]{@{}c@{}}0.0418\\(0.0150)\end{tabular} & \begin{tabular}[c]{@{}c@{}}0.0491\\(0.0182)\end{tabular} & \begin{tabular}[c]{@{}c@{}}0.0348\\(0.0143)\end{tabular} & \begin{tabular}[c]{@{}c@{}}\textbf{0.0334}\\\textbf{(0.0131)}\end{tabular}\\
Substantia nigra & \begin{tabular}[c]{@{}c@{}}0.0651\\(0.0233)\end{tabular} & \begin{tabular}[c]{@{}c@{}}0.0658\\(0.0262)\end{tabular} & \begin{tabular}[c]{@{}c@{}}0.0802\\(0.0332)\end{tabular} & \begin{tabular}[c]{@{}c@{}}0.0555\\(0.0172)\end{tabular} & \begin{tabular}[c]{@{}c@{}}\textbf{0.0519}\\\textbf{(0.0168)}\end{tabular}\\
\bottomrule
\end{tabular}
\end{table}

CONCERT uses individual-level source data and selects more variables on average. TRACE instead uses 
source coefficient summaries and effective sample sizes, allowing its application when individual-level
source records cannot be shared.

\section{Conclusion}
\label{sec:conclusion}
We proposed TRACE, a Bayesian transfer-learning method for variable selection using fitted coefficients 
from multiple source studies. TRACE learns a weighted combination of source estimates and uses separate 
indicators for target inclusion and coordinate-specific borrowing. This allows a target variable to be 
selected even when the combined source information is not transferable. An approximated collapsed variational 
EM algorithm jointly estimates these components, with robust initialization and convex source-weight updates.

In fixed dimension, we establish consistency of target selection, transfer decisions, and source-weight estimation 
for any prespecified finite number of iterations. In growing dimension, we establish consistency of the source weights 
and joint recovery of the target signal set and transfer configuration on the active coordinates at global optima of 
the exact collapsed variational objective.

Simulations show that TRACE achieves competitive variable-selection performance under different forms of source heterogeneity. Its gains
are particularly apparent with complementary sources, where TRACE-EM closely matches Oracle TRACE. In the GTEx application, TRACE achieves
lower mean prediction error than target-only ACVB in five of six tissues under a common post-selection refit.

TRACE has several limitations that suggest directions for future research. First, the method treats fitted source 
coefficients as fixed inputs and uses source sample sizes to determine the strength of borrowing. It does not 
explicitly account for the estimation uncertainty or dependence among source coefficient estimates. 
Incorporating standard errors or selected covariance summaries could improve the calibration of source information. 
However, reliable estimation of these quantities can be difficult in high dimensions. Second, TRACE uses source weights 
shared across predictors. Although the transfer indicators allow borrowing to be rejected for individual variables, 
they do not allow different variables to use different source combinations. Structured or group-specific source weights 
could provide greater flexibility, but would require additional regularization to maintain stable estimation from 
limited source summaries. The present framework is developed for Gaussian linear regression. Extensions to generalized 
linear models would require new approximations to the collapsed likelihood and corresponding theoretical analysis. 
Another important direction concerns the practical variational algorithm. Our fixed-dimensional results apply to 
any prespecified finite number of iterations, whereas the high-dimensional results concern global optima of the exact 
collapsed variational objective. Establishing convergence properties and assessing approximation errors for the implemented 
algorithm in growing dimensions remain open problems.

\section*{Funding}
This study was funded by the National Natural Science Foundation of China (Grant No. 12401377).

\section*{Disclosure Statement}
The authors declare no conflicts of interest.

\section*{Declaration of Generative AI Use}
During the preparation of this manuscript, the authors utilized ChatGPT and OpenAI Codex (GPT-6 and GPT-6.1, OpenAI) to assist with language refinement, theoretical derivations, and computational code. 
All literature citations and references were independently retrieved and verified by the authors 
from original sources. The authors thoroughly reviewed, validated, and take full responsibility 
for all theoretical mathematical proofs, code implementation, numerical results, and the overall 
scientific integrity. The authors have verified the originality and 
accuracy of all AI-assisted content and reviewed the terms of use of the AI tools and 
confirmed that this content is suitable for publication under those terms.

\section*{Supplementary Material} 
The online supplementary material includes additional theoretical results, proofs, details of the 
variational algorithm, and additional results of simulations and applications.

\bibliographystyle{trace_jasa}
\bibliography{reference}

\end{document}